\documentstyle[aps,epsf,rotate,multicol]{revtex}

\begin{document}

\draft

\title{Small-world networks: Evidence for a crossover picture}

\author{Marc Barth\'el\'emy{\footnote{Permanent address: CEA-BIII, 
Service de Physique de la Mati\`ere Condens\'ee, France.}}
and Lu\'{\i}s A. Nunes Amaral}

\address{Center for Polymer Studies and Dept. of Physics,
		Boston University, Boston, MA 02215 }


\maketitle

\begin{abstract}

Watts and Strogatz [Nature {\bf 393}, 440 (1998)] have recently
introduced a model for disordered networks and reported that, even for
very small values of the disorder $p$ in the links, the network behaves
as a small-world.  Here, we test the hypothesis that the appearance of
small-world behavior is not a phase-transition but a crossover
phenomenon which depends both on the network size $n$ and on the
degree of disorder $p$.  We propose that the average distance $\ell$
between any two vertices of the network is a scaling function of
$n/n^*$. The crossover size $n^*$ above which the network behaves as a
small-world is shown to scale as $n^*(p \ll 1) \sim p^{-\tau}$ with
$\tau \approx 2/3$.
\end{abstract}

\begin{multicols}{2}


Two limiting-case topologies have been extensively considered in the
literature.  The first is the regular lattice, or regular network,
which has been the chosen topology of innumerable physical models such
as the Ising model or percolation\cite{Stanley71,Stanley90,Bunde96}.
The second is the random graph, or random network, which has been
studied in mathematics and used in both natural and social sciences
\cite{Winfree80,Kuramoto84,Gerhardt90,Nowak92,Wasserman94,
Braiman95,Collins95,Hopfield95,Crutchfield95,Wiesenfeld96,Hess96,
Kretschmar96,Bressloff97}.

Erd\"os and co-workers studied extensively the properties of random
networks ---see\cite{Bollobas85} for a review.  Most of this work
concentrated on the case in which the number of vertices is kept
constant but the total number of links between vertices
increases\cite{Bollobas85}: The Erd\"os-R\'enyi result \cite{Erdos59}
states that for many important quantities there is a percolation-like
transition at a specific value of the average number of links per
vertex.  In physics, random networks are used, for example, in studies
of dynamical problems\cite{Christensen98,Barrat98}, spin models and
thermodynamics\cite{Barrat98,Luque97}, random walks\cite{Cassi96}, and
quantum chaos\cite{Kottos97}.  Random networks are also widely used in
economics and other social sciences\cite{Wasserman94,Axelrod84,Jain98}
to model, for example, interacting agents.

In contrast to these two limiting topologies, empirical
evidence\cite{Watts98,Collins98} suggests that many biological,
technological or social networks appear to be somewhere in between
these extremes.  Specifically, many real networks seem to share with
regular networks the concept of neighborhood, which means that if
vertices $i$ and $j$ are neighbors then they will have many common
neighbors ---which is obviously not true for a random network.  On the
other hand, studies on epidemics\cite{Hess96,Kretschmar96,Watts98}
show that it can take only a few ``steps'' on the network to reach a
given vertex from any other vertex.  This is the foremost property of
random networks, which is not fulfilled by regular networks.

\begin{figure}
\vspace*{-2.0cm}
\centerline{
\epsfysize=1.4\columnwidth{\epsfbox{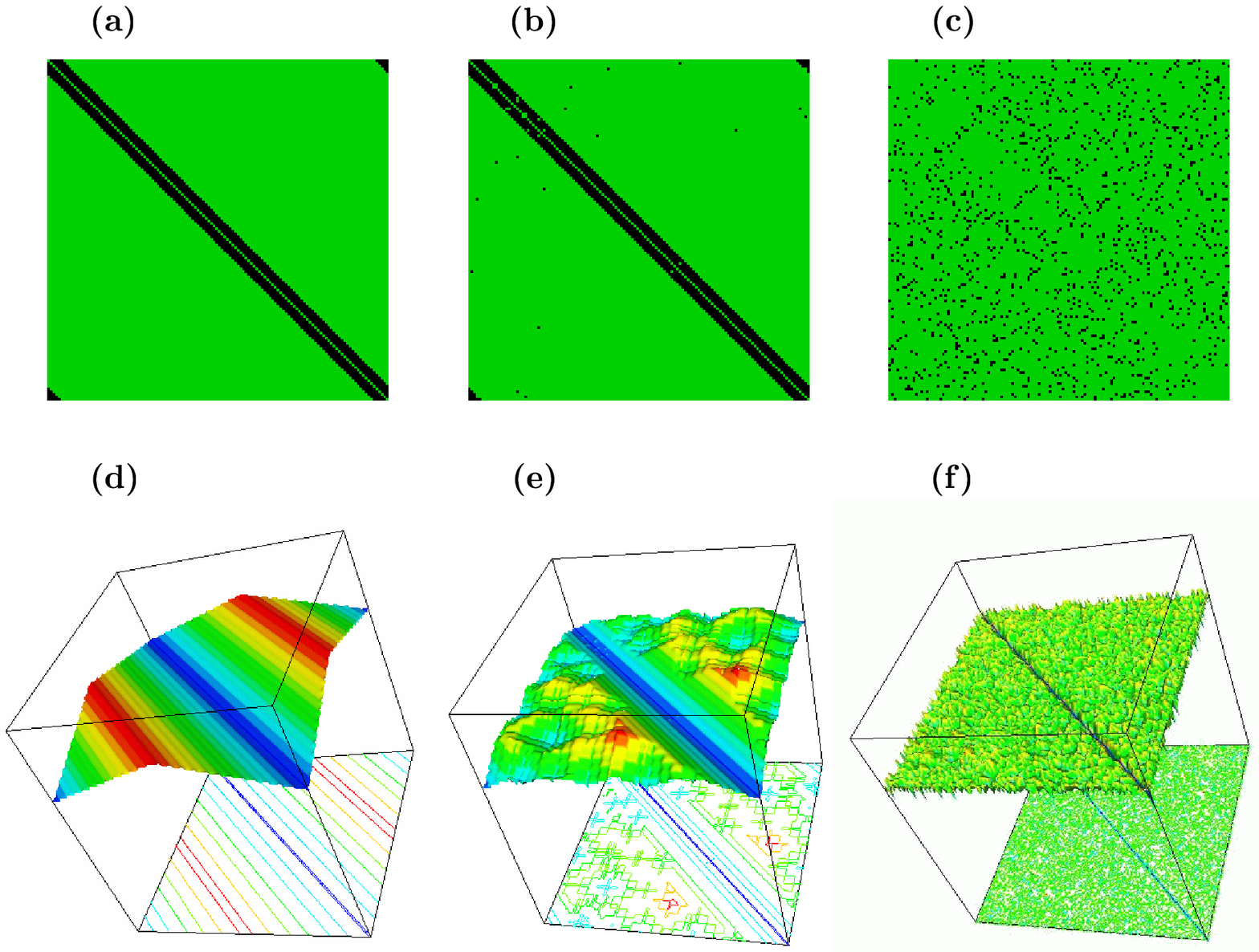}}
}
\vspace*{-8.0cm}
\caption{ Effect of disorder on the distance between vertices of the
network (go to http://polymer.bu.edu/$\sim$amaral/Networks.html for
color pictures). We consider here matrices with $z=10$, $n=128$ and
with periodic boundary conditions, that is, vertex 1 follows vertex
$n$. Adjacency matrices for (a) a regular one-dimensional network
where each vertex is connected to its $z$ nearest neighbors, (b) a
disordered network with $p = 0.01$, and (c) a random network.  Black
indicates that a link is present between the two vertices while gray
(green in the color picture) indicates the absence of a link.  Note
that (a) and (b) are nearly identical.  Distance matrices for (d) the
regular network, (e) the disordered network with $p = 0.01$, and (f)
the random network.  We use the relief of the surface and a color
scheme to represent the distance between two vertices.  Greater height
indicates larger distance. The color scheme is the same for the relief
and for the contour lines: Distance increases from very dark gray
(blue) to gray (green) to light gray (yellow) to dark gray (red).  For
the regular network, the contour lines are parallel to the
diagonal. On the other hand, for the disordered network the contour
lines ``circle'' around specific links that act as ``through-ways'' of
the network. This effect prevents the distance between any two
vertices from ever becoming large, that is, of the order of the system
size.}
\label{f.picture}
\end{figure}

To bridge the two limiting cases, and to provide a model for
real-world systems\cite{Milgram67,Kochen89}, Watts and
Strogatz\cite{Watts98,Collins98} have recently introduced a new type
of network which is obtained by randomizing a fraction $p$ of the
links of the regular network. As Ref.~\cite{Watts98}, we consider as
an initial structure ($p=0$) the one-dimensional regular network where
each vertex is connected to its $z$ nearest neighbors. For $0 < p <
1$, we denote these networks disordered, and keep the name random
network for the case $p=1$.  Reference~\cite{Watts98} reports that for
a small value of the parameter $p$ ---which interpolates between the
regular ($p=0$) and random ($p=1$) networks--- there is an onset of
small-world behavior. The small-world behavior is characterized by the
fact that the distance between any two vertices is of the order of
that for a random network and, at the same time, the concept of
neighborhood is preserved, as for regular lattices
[Fig.\ref{f.picture}].  The effect of a change in $p$ is extremely
nonlinear as is visually demonstrated by the difference between
Figs.~\ref{f.picture}a,d and Figs.~\ref{f.picture}b,e where a very
small change in the adjacency matrix leads to a dramatic
change in the distance between different pairs of vertices.


Here, we study the origins of the small-world
behavior\cite{Milgram67,Kochen89}.  In particular, we investigate if
the onset of small-world networks is a phase transition or a crossover
phenomena.  To answer this question we consider not only changes in
the value of $p$ but also in the system size $n$.

The motivation for this study is the following.  In a regular
one-dimensional network with $n$ vertices and $z$ links per vertex, the
average distance $\ell$ between two vertices increases as $n/(2z)$
---the distance is defined as the minimum number of steps between the
two vertices.  The regular network is similar to Manhattan: Walking
along $5^{th}$ Avenue from Washington Square Park in $4^{th}$ Street to
Central Park in $59^{th}$ Street, we have to go past 55 blocks.  On the
other hand, for a random network, each ``block'' brings us to a point
with $z$ new neighbors.  Hence, the number of vertices increases with
the number of steps $k$ as $n \sim z^{k}$, which implies that $\ell$
increase as $\ln n/\ln z$.  The random network is then like a strange
subway system that would directly connect different parts of Manhattan
and enable us to go from Washington Square Park to Central Park in just
one stop.  In view of these facts, it is natural to enquire if the
change from large-world ($\ell \sim n$) to small-world ($\ell \sim \ln
n$) in disordered networks occurs through a phase transition for some
given value of $p$\cite{transition} or if, for any value of $p$, there
is a crossover size $n^*(p)$ below which our network is a large-world
and above which it is a small-world.

In the present Letter we report that the appearance of the small-world
behavior is not a phase-transition but a crossover phenomena.  We
propose the scaling ansatz
\begin{equation}
\ell(n,p) \sim n^* F\left( \frac{n}{n^*} \right)\,,
\label{e.scaling}
\end{equation}
where $F(u \ll 1) \sim u$, and $F( u \gg 1) \sim \ln u$, and $n^*$ is a
function of $p$\cite{note_on_z}. Naively, we would expect that when the
average number of rewired links, $pnz/2$, is much less than one, the
network should be in the large-world regime. On the other hand, when
$pnz/2\gg 1$, the network should be a small-world\cite{Herzel98}. Hence,
the crossover size should occur for $n^* p = O(1)$, which implies $n^*
\sim p^{-\tau}$ with $\tau = 1$. This result relies on
the fact that the crossover from large to small worlds is obtained with
only a small but finite fraction of rewired links.  We find that the
scaling ansatz (\ref{e.scaling}) is indeed verified by the average
distance $\ell$ between any two vertices of the network. We also
identify the crossover size $n^*$ above which the network behaves as a
small-world, and find that it scales as $n^*\sim p^{-\tau}$ with $\tau
\approx 2/3$, distinct from the trivial expectation $\tau = 1$.


Next, we define the model and present our results.  We start from a
regular one-dimensional network with $n$ vertices, each connected to $z$
neighbors.  We then apply the ``rewiring'' algorithm of \cite{Watts98}
to this network.  The algorithm prescribes that every link has a
probability $p$ of being broken and replaced by a new random link.  We
replace the broken link by a new one connecting one of the original
vertices to a new randomly selected vertex.  Each of the other $n-2$
vertices ---we exclude the other vertex of the broken link--- has an
{\it a priori\/} equal probability of being selected but we then make
sure that there are no duplicate links.  Hence, the algorithm preserves
the total number of links which is equal to $nz/2$. A quantity that is
affected by the rewiring algorithm is the probability distribution of
local connectivities.  For $p\simeq 0$, this probability is narrowly
peaked around $z$, but it gets broader with increasing $p$.  For $p=1$,
the average and the standard deviation of the local connectivity are of
the same order of magnitude and equal to $z$.

Once the disordered network is created, we calculate the distance
between any two vertices of the network and its average value $\ell$.
To calculate all the distances between vertices, we use the
Moore-Dijkstra algorithm\cite{Gondran84} whose execution time scales
with network size as $n^3 \ln n$.  We perform between 100 and 300
averages over realizations of the disorder for each pair of values of
$n$ and $p$. 

\begin{figure}
\narrowtext
\vspace*{0.0cm}
\centerline{
\epsfysize=0.9\columnwidth{\rotate[r]{\epsfbox{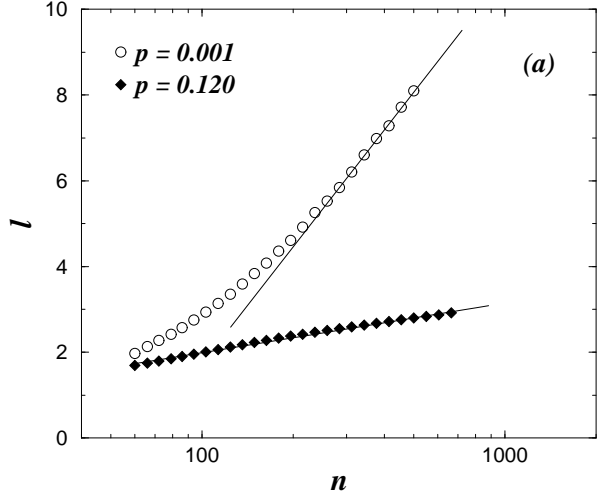}}}
}
\vspace*{0.3cm}
\centerline{
\epsfysize=0.9\columnwidth{\rotate[r]{\epsfbox{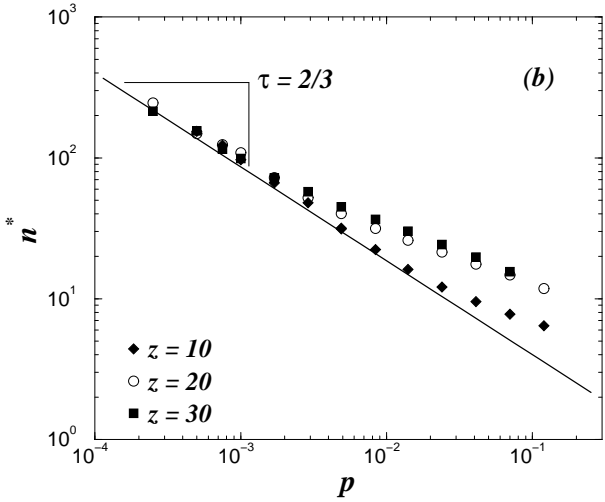}}}
}
\vspace*{0.0cm}
\caption{ Determination of the crossover size $n^*$. (a) Semi-log plot
of $\ell$ versus network size for two representative values of $p$ and
for $z=20$.  Following
Eqs.~(\protect\ref{e.scaling}-\protect\ref{e.tau}), we can determine
$n^*$ ---apart from a multiplicative constant--- from the asymptotic
slope of $\ell$ against $\ln n$.  (b) Scaling of $n^*$ with $p$ for the
three values of $z$ discussed in the text. The curves for $z=20$ and
$30$ have been shifted up so as to coincide in the region where they
scale as a power law.  Following Eq.~(\protect\ref{e.tau}), we make a
power law fit to $n^*(p)$ for $p \ll 1$ and obtain $\tau \approx 2/3$. }
\label{f.ncross}
\end{figure}

Here, we present results for three values of connectivity $z =10, 20$
and $30$ and system sizes up to $1000$.  The scaling ansatz
(\ref{e.scaling}) enables us to determine $n^*(p)$ from $\ell(n)$ at
fixed $p$.  Indeed, $\ell(n \gg n^*) \sim n^* \ln n$ which implies
that $n^*$ is the asymptotic value of $d\ell/d(\ln n)$
[Fig.~\ref{f.ncross}(a)]. Figure~\ref{f.ncross}(b) shows the
dependence of $n^*$ on $p$ for different values $z$.  We hypothesize
that
\begin{equation}
n^* - \frac{1}{\ln z} \sim p^{-\tau} g(p)\,, 
\label{e.ncross}
\end{equation}
where the term in $z$ arises from the fact that $\ell = \ln n / \ln z$
for a random network ($p=1$), and $g(p \to 1) \to 0$.  Moreover, 
$g(p)$ approaches a constant as $p \to 0$, leading to
\begin{equation}
n^*\sim p^{-\tau}\,,
\label{e.tau}
\end{equation}
for small $p$. Due to the effect of $g$ and the fact that $n < 1000$ in
our numerical simulations, we are constrained to estimate $\tau$ from
the region $2.5\times10^{-4} < p < 2\times10^{-2}$.  For all values of
$z$ we obtain $\tau = 0.67 \pm 0.10$ [Fig.~\ref{f.ncross}].

\begin{figure}
\narrowtext
\vspace*{0.0cm}
\centerline{
\epsfysize=0.9\columnwidth{\rotate[r]{\epsfbox{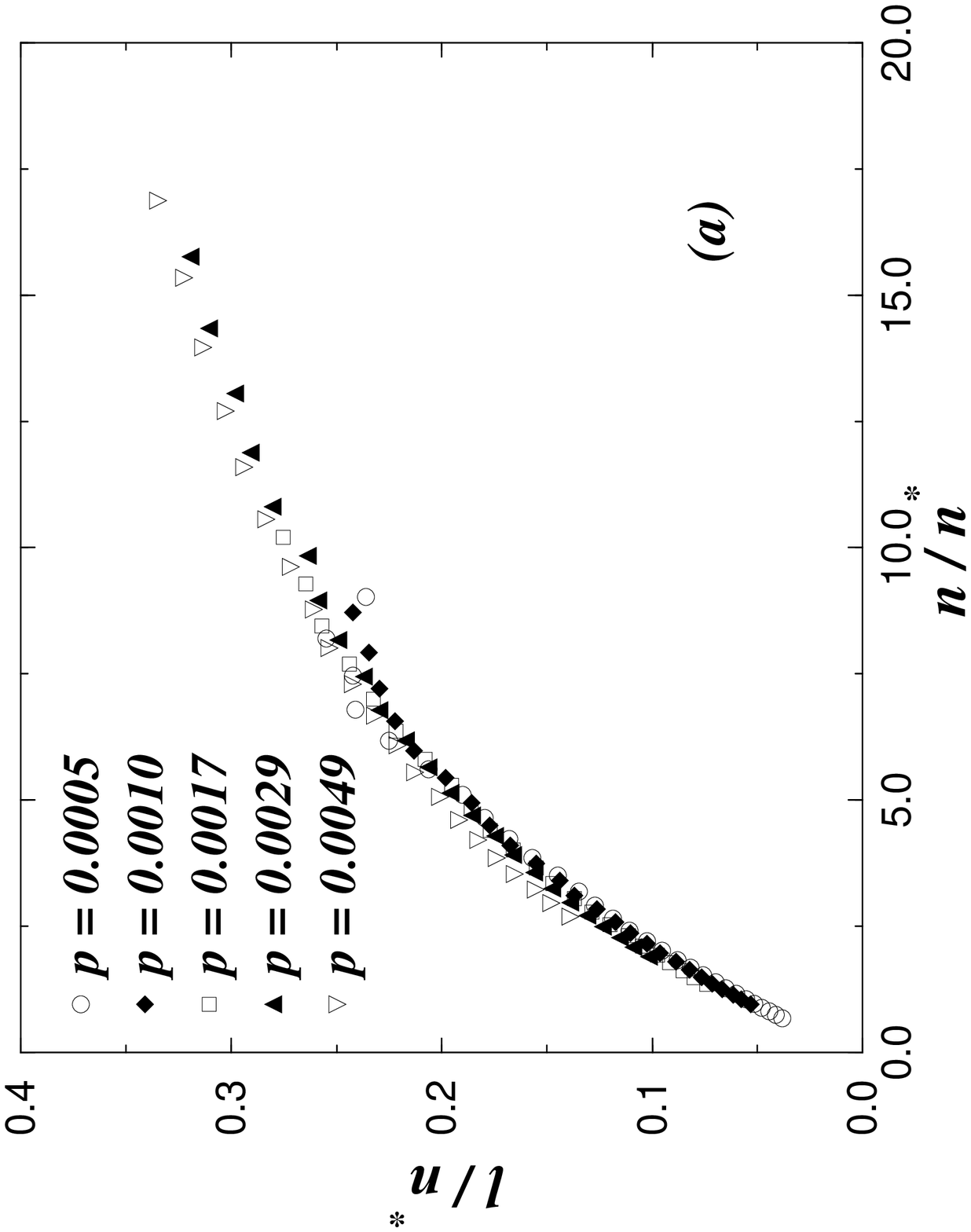}}}
}
\vspace*{0.0cm}
\centerline{
\epsfysize=0.9\columnwidth{\rotate[r]{\epsfbox{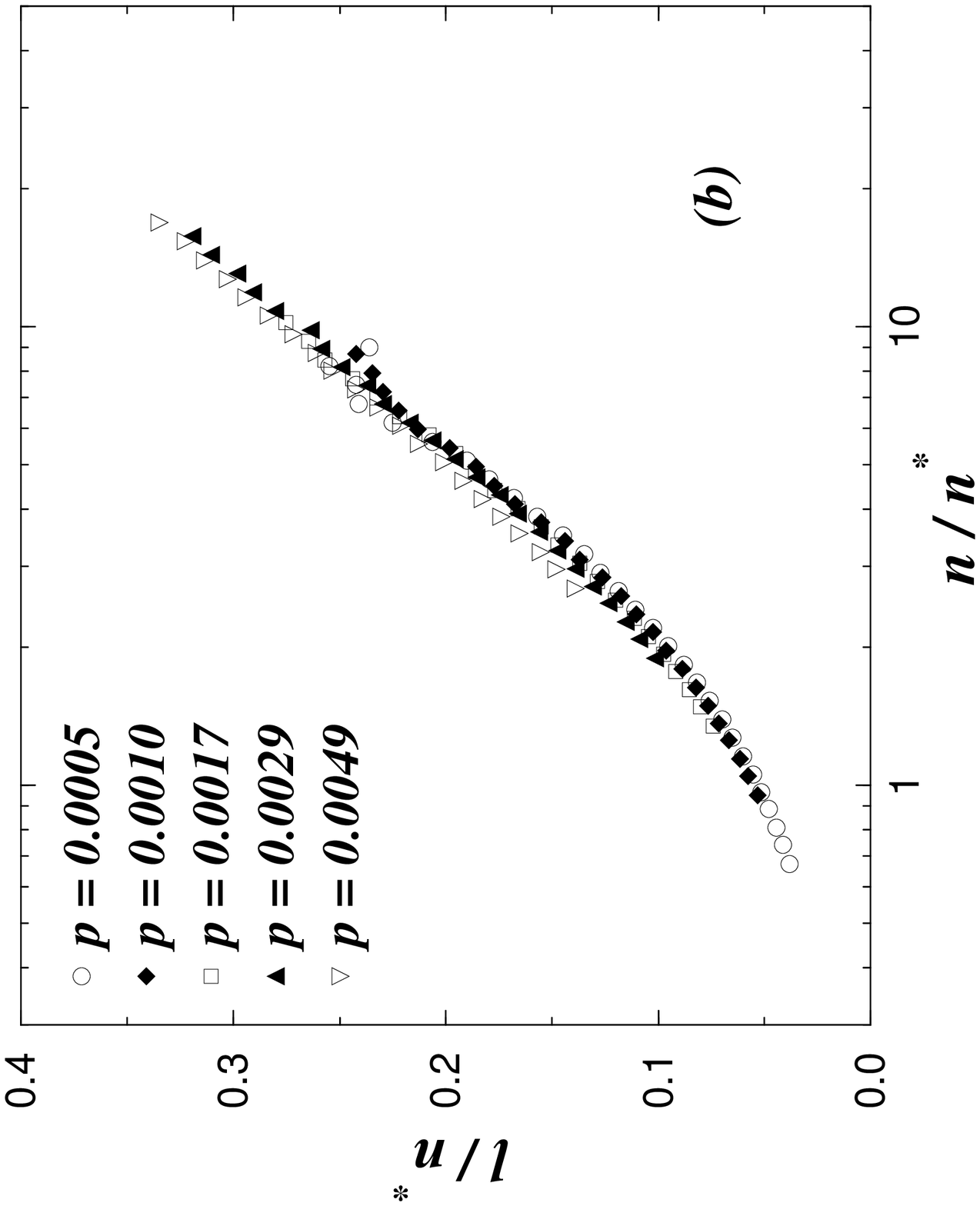}}}
}
\vspace*{0.0cm}
\caption{ Data collapse of $\ell(n,p)$ for $z=10$ and different values
of $p$ and $n$.  (a) Plot of the scaled average distance between
vertices $\ell / n^*$ versus scaled system size $n/n^*$. (b) Same data
as in (a) but in a semi-log plot.  Note the linear behavior of the data
for $n < n^*$ and the logarithmic increase of $\ell$ for large system
sizes.  }
\label{f.collapse}
\end{figure}

Using this value of $\tau$ and the scaling form (\ref{e.scaling}) we are
able to collapse all the values of $\ell(n,p)$ onto a single curve
[Fig.~\ref{f.collapse}].  This data collapse confirms our scaling ansatz
and estimate of $\tau$.


In summary, we have shown that the onset of small-world behavior is a
crossover phenomena and not a phase transition from a large-world to a
small one. The crossover size scales as $p^{-\tau}$ with $\tau\simeq
2/3$. The surprising fact that $\tau < 1$, shows that the rewiring
process is highly nonlinear and can have dramatic consequences on the
global behavior of the network.  This implies that in order to {\it
decrease\/} the radius of a network it is necessary to rewire only a few
links. We also note that the value of the exponent $\tau$ will likely
depend on the dimensionality of the initial regular network. This point
will be adressed in future work.

We believe that the disordered networks introduced in \cite{Watts98}
may constitute a promising topology for more realistic studies of many
important problems such as flow in electric-power or information
networks, spread of epidemics, or financial systems.  The results
reported here support this hypothesis because they suggest that, for
{\it any\/} given degree of disorder of the network, if the system is
larger than the crossover size, the network will be in the small-world
regime.

We thank S.V.~Buldyrev, L.~Cruz, P. Gopikrishnan, P.~Ivanoc,
H.~Kallabis, E.~La~Nave, T.J.P.~Penna, A.~Scala, and H.E.~Stanley for
stimulating discussions.  L.A.N.A.\ thanks the FCT/Portugal and M.B.\
thanks the DGA for financial support.



\end{multicols}

\vfill
\eject

\begin{multicols}{2}

\end{multicols}


\begin{references}
\bibitem{transition}
Reference~\protect\cite{Watts98} describes a ``onset of the small-world
phenomena'' which suggests a transition between different ``phases.''
This impression is reinforced by the fact that in
\protect\cite{Watts98} there is no mention of a dependence of the
results on network size.
\bibitem{note_on_z}
Note that there are factors depending on $z$ which we will ignore here.
\end{references}
\end{document}